\documentclass[aps,twocolumn]{revtex4}
\usepackage{dcolumn}
\usepackage{graphicx}
\usepackage{amsmath}
\usepackage{amsfonts}
\usepackage{amssymb}
\usepackage{psfrag}
\usepackage{wrapfig}
\usepackage{subfigure}
\usepackage{makeidx}
\usepackage{bm}
\usepackage{epsf}
\usepackage[colorlinks,urlcolor=blue,citecolor=blue]{hyperref}

\begin{document}
\title{Note on Dirac monopole theory and Berry geometric phase}
\author{Li-Chen Zhao} \email{zhaolichen3@nwu.edu.cn}
\address{$^{1}$School of Physics, Northwest University, Xi'an 710127, China}
\address{$^{2}$Peng Huanwu Center for Fundamental Theory, Xi'an 710127, China}
\address{$^{3}$Shaanxi Key Laboratory for Theoretical Physics Frontiers, Xi'an 710127, China}

\begin{abstract}
This work reveals the intrinsic connection between Dirac monopole theory and Berry geometric phases by extending Dirac's theory to the parameter space. Using the simplest two-mode Hamiltonian model, we explicitly visualize Dirac strings with endpoints in the parameter space, demonstrating that these endpoints correspond to accidental degeneracy points of energy eigenvalues in Hermitian systems. We show that non-integrable phase factors, induced by such Dirac strings, directly give rise to the well-known Berry connection and curvature, which can be derived rigorously via Dirac's monopole framework. Our results indicate that the Berry geometric phase is essentially the non-integrable phase factor induced by Dirac strings with endpoints in the parameter space. This establishes a unified and effective approach to study monopoles and geometric phases, particularly applicable when Berry's framework fails.
\end{abstract}
\date{\today}
\maketitle

\section{Introduction}
In 1931, Dirac first proposed the concept of virtual monopoles associated with nodal singularities (Dirac strings) of wavefunctions, revealing that the non-commutative property of phase gradients near these singularities gives rise to non-integrable phase factors \cite{Dirac}. A critical feature of this theory is the endpoint of the Dirac string, which acts as the monopole itself. However, the visualization of such Dirac strings with endpoints has long remained challenging: although explicit examples of Dirac monopoles have been demonstrated in real space \cite{Ferrell,Ray,XFZhou,XFZhou2,JHZheng}, the corresponding physical models or operational processes are overly complex, making it difficult for non-experts to grasp the generality and elegance of Dirac's monopole theory.

In 1984, Berry systematically established the geometric phase theory, showing that accidental degeneracy points of energy eigenvalues can be analogous to virtual monopoles \cite{Berry1}. This theory successfully explains striking phase phenomena observed in various physical systems \cite{Ryt,Vlad,Panch,Herzberg,Longuet,Smith,Burdden,Yang}. Despite noting similarities between the Berry curvature and Dirac strings, the intrinsic connection between Dirac's monopole theory and Berry's geometric phase theory remains unclear. A key obstacle is that normalized eigenstates---commonly used in standard treatments---obscure the density zeros of wavefunctions, thereby hiding Dirac strings. Additionally, the role of non-integrable phase factors in this context has not been fully clarified. These issues have led the scientific community to predominantly adopt Berry's framework \cite{RMP1,band2,RMP2} rather than Dirac's original perspective. Furthermore, Berry's framework is constrained by adiabatic conditions, limiting its generality, whereas Dirac's theory holds promise for a more universal description \cite{Aharonov1987,Zhaoliu1,Zhaoliu2}.

This work aims to clarify the intrinsic relationship between Dirac's monopole theory \cite{Dirac} and Berry's geometric phase theory \cite{Berry1} by extending Dirac's monopole theory from real space to the parameter space. Using the simplest two-mode Hamiltonian, we explicitly visualize Dirac strings and their endpoints, demonstrating that these endpoints correspond to energy degeneracy points in Hermitian systems. We derive the Berry connection and curvature directly from the non-integrable phase factors in Dirac's theory, thus establishing that Berry geometric phases originate from Dirac strings with endpoints. This extension of Dirac's theory is non-trivial: in the parameter space, different eigenstates exhibit distinct monopole charges, a feature absent in Dirac's original real-space monopole theory \cite{Dirac}. We further explicitly demonstrate the existence of Dirac strings with endpoints and calculate the phase variations of closed curves around the Dirac string, its endpoint, and other points---results that facilitate a deeper understanding of Dirac's original descriptions of monopoles \cite{Dirac}. These findings are expected to inspire further studies on geometric phases using or extending Dirac's monopole theory, particularly in scenarios where Berry's framework is inapplicable.
\begin{figure}[t]
\begin{center}
\includegraphics[width=\linewidth]{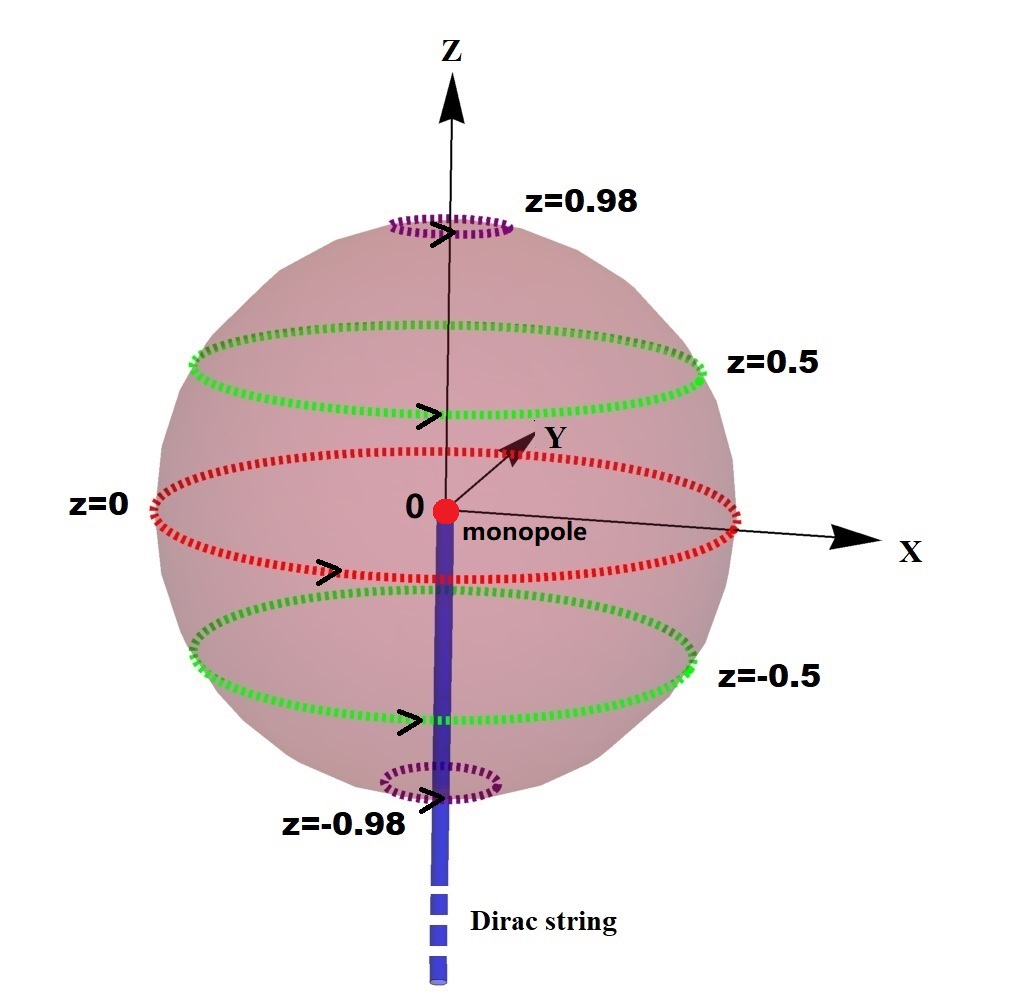}
\end{center}
\caption{The Dirac string and its endpoint at $(0,0,0)$ for  the eigenstate of $E_{+}(\mathbf{R})$ branch. Five closed curves around the Dirac string, its endpoint, and other points, are chosen to calculate the phase variations.}\label{Fig1}
\end{figure}

\section{Dirac string and non-integrable phase factor}
To visualize Dirac strings, their endpoints, and the underlying geometric phases, we adopt one of the simplest two-mode Hamiltonians \cite{Berry1}, given by:
\begin{eqnarray}\label{Hamiltonian}
H(\mathbf{R})=
\left(
\begin{array}{cc}
	Z &X-i Y \\
	X+i Y & -Z \\
	\end{array}
\right),
\end{eqnarray}
where \(X, Y, Z\) are three independent parameters, and \(\mathbf{R} = (X, Y, Z)\) denotes a vector in the parameter space. The energy eigenvalues of \(H(\mathbf{R})\) are \(E_{\pm}(\mathbf{R}) = \pm \sqrt{X^2 + Y^2 + Z^2} = \pm R\), with the degeneracy point of the energy spectrum located at \((0, 0, 0)\) in the parameter space. The eigenstates corresponding to these two branches are:
\begin{eqnarray}\label{positiveV}
|V_{\pm}(\mathbf{R})\rangle = \left(
\begin{array}{c}
Z \pm R \
X + iY \
\end{array}
\right).
\end{eqnarray}
In standard quantum mechanics textbooks \cite{Schiff}, the phase of eigenstates is typically set to zero by convention, resulting in no explicit phase information in the above eigenstates. However, such conventional phase assignments break down when non-integrable phase characteristics emerge in the parameter space. Although the non-integrable phase factor is not explicitly present in the eigenvectors above, they can still be used to calculate the Berry curvature in the presence of geometric phases \cite{Berry1}. A virtual monopole has been suggested to reside at the accidental degeneracy point of energies, yet the Dirac monopole picture remains obscure even in this scenario. Thus, we aim to reconstruct Dirac's original framework here to clarify the intrinsic relationship between Dirac's monopole theory \cite{Dirac} and Berry's geometric phases \cite{Berry1}.In the context of Dirac's monopole theory \cite{Dirac}, the first step is to examine the zeros of wavefunctions, which form nodal lines (commonly referred to as Dirac strings). The density zeros of \(|V_{+}(\mathbf{R})\rangle\) are illustrated in Fig.~\ref{Fig1}, revealing a nodal line along the negative z-direction for the \(|V_{+}(\mathbf{R})\rangle\) state, with this nodal line terminating at \((0, 0, 0)\). According to Dirac's monopole theory, the relative phase between two points in the parameter space depends on the integration path of the phase gradient field, giving rise to the so-called non-integrable phase factor. This behavior stems from phase singularities at the wavefunction zeros. To elaborate, consider an arbitrary complex wavefunction \(\psi = \text{Re}(\psi) + i\text{Im}(\psi)\), whose phase can be expressed as: \(\phi = \arctan\left(\frac{\text{Im}(\psi)}{\text{Re}(\psi)}\right) + n\pi,\)where $n$ is determined by the direction of the phase gradient. Evidently, wavefunction zeros correspond to phase singularities. We analyze the phase in the abstract parameter space \((\text{Re}(\psi), \text{Im}(\psi))\), where the vector potential associated with the phase is defined as:
\begin{eqnarray}\label{field}
\bm{A}&=&\frac{\partial\phi }{\partial Re(\psi)}\bm{e_{Re}}+\frac{\partial \phi}{\partial Im(\psi)}\bm{e_{Im}} \nonumber\\
&=&-\frac{Im(\psi)}{Re(\psi)^2+Im(\psi)^2}\bm{e_{Re}}+\frac{Re(\psi)}{Re(\psi)^2+Im(\psi)^2}\bm{e_{Im}}. \nonumber \\
\end{eqnarray}
Wavefunction zeros naturally correspond to singularities of the vector potential \(\bm{A}\). For closed line integrals of \(\bm{A}\) around such singularities, the result is \(\pm 2\pi\), marking the presence of non-integrable phase factors in this abstract space. The topological charge in the \((\text{Re}(\psi), \text{Im}(\psi))\) space is \(\pm 1\), and it may vary when mapping the wavefunction to specific real or parameter spaces. This explains why wavefunction zeros are fundamentally critical to topological effects across diverse spaces \cite{Zhaoliu1,Zhaoliu2,DuanPRE1999,DuanPRE199960,HePRE2021,Yu2024}. We further investigate whether nodal lines possess endpoints. If a nodal line has no endpoints, the phase difference between any two paths is always \(2m\pi\) (where $m$is an integer), with no observable experimental consequences---rendering the introduction of geometric phases unnecessary. In contrast, if a nodal line terminates at a finite endpoint, phase differences between paths will deviate from \(2m\pi\) and become experimentally observable. Applying Stokes' theorem to a closed surface, the endpoint of a Dirac string is predicted to correspond to a monopole with charge \(\mu = m/2\) \cite{Dirac}. Notably, Dirac's original description of nodal lines and their endpoints was based on an abstract wavefunction \cite{Dirac}. Subsequent efforts have sought explicit wavefunctions exhibiting Dirac monopoles \cite{Ferrell,Ray}, but these examples are overly complex for non-experts. To our knowledge, this is the first demonstration of a wavefunction with a Dirac string and its endpoint using the simplest two-mode model. It is emphasized that the eigenstate remains unnormalized here: normalization would hide density zeros in the \((X, Y, Z)\) parameter space, explaining why Dirac strings are difficult to visualize in the parameter space using normalized eigenstates \cite{Berry1}. While Dirac's original analysis focused on wavefunctions in real space and time, the eigenstates discussed herein reside in the parameter space---nevertheless, their phase variations can be analyzed analogously to those in real space.
\begin{figure}[t]
\begin{center}
\includegraphics[width=\linewidth]{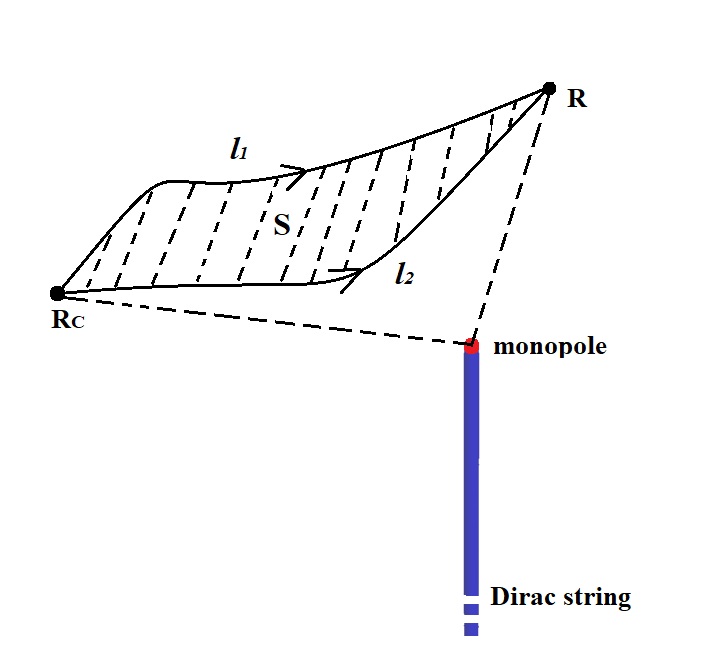}
\end{center}
\caption{The illustration of non-integrable phase factor, predicted by the existence of monopole and Dirac string. The relative phase between $\mathbf{R}_c$ and $\mathbf{R}$ depends on the integration pathes.}\label{Fig0}
\end{figure}

The existence of endpoints of Dirac strings qualitatively implies that the phases of the aforementioned eigenstates exhibit non-integrable characteristics and possess observable effects \cite{Dirac}. We introduce a transformation of the eigenstates as \(|V_{\pm}(\mathbf{R})\rangle \rightarrow |V_{\pm}(\mathbf{R})\rangle e^{i \gamma_{\pm}}\), where \(\gamma_{\pm}\) denotes the non-integrable phase factor. Under this transformation, the eigenvalue equation remains valid:
\begin{eqnarray}\label{rightvector}
H(\mathbf{R})|V_{\pm}(\mathbf{R})\rangle e^{i \gamma_{\pm}} =E_{\pm}(\mathbf{R})|V_{\pm}(\mathbf{R})\rangle e^{i \gamma_{\pm}}.
\end{eqnarray}
While the eigenvalue equation itself does not provide direct information about \(\gamma_{\pm}\), the phase factor necessarily exists when Dirac strings have endpoints at finite locations. This qualitatively explains why the non-integrable phase factor was introduced in Eq. (3) of Berry's original work \cite{Berry1}. Importantly, the presence of non-integrable phase factors does not alter the properties of Dirac strings or their endpoints. Even eigenstates without an explicit non-integrable phase factor (as in Eq.~\eqref{positiveV}) can still reveal Dirac strings with endpoints, which qualitatively predict the existence of such phase factors. In the absence of Dirac strings with endpoints, setting \(\gamma_{\pm}=0\) remains valid, and conventional phase conventions hold---providing a reliable criterion to judge the presence of non-trivial non-integrable phase factors based on the nodal line properties of eigenstates.

We further analyze the properties of the non-integrable phase factor \(\gamma_{\pm}\) using Dirac's original monopole theory \cite{Dirac}. Due to the existence of nodal lines and their endpoints, the relative phase between any two points in the parameter space depends on the integration path of the phase gradient, as illustrated in Fig.~\ref{Fig0}. Specifically, the relative phase between \(\mathbf{R}_c\) and \(\mathbf{R}\) calculated along two different paths \(\textbf{\emph{l}}_1\) and \(\textbf{\emph{l}}_2\) generally yields distinct results. This path dependence is the defining feature of a non-integrable phase. Notably, the difference between these phase values---corresponding to a closed path formed by the two paths---can be directly determined by the magnetic flux of the monopole field and the contribution from Dirac strings through the surface \(\mathbf{S}\) bounded by the closed path, via Stokes' theorem \cite{Dirac}. The phase variation around any closed path in the parameter space is given by:
\begin{eqnarray}
\Delta \gamma_{\pm} = \iint \textbf{B}(\mathbf{R}) \cdot d \textbf{S}+ 2 \pi \sum m, \label{eq15}
\end{eqnarray}
where \(\textbf{B}(\mathbf{R})= \nabla \times \textbf{A}(\mathbf{R})=\mu \mathbf{R}/R^3\) in isotropic space, and \(\sum m\) represents the contribution from Dirac strings, with each $m$ being an integer determined by the properties of the respective string. This property directly indicates that the non-integrable phase factor \(\gamma_{\pm}\) depends solely on the parameter \(\mathbf{R}\), i.e., \(\gamma_{\pm}\) should be expressed as \(\gamma_{\pm}(\mathbf{R})\). If the charges of the strings and magnetic monopoles are known, phase variations can be directly predicted using Eq.~\eqref{eq15}. This intuitive picture clarifies why Berry chose to investigate phase variations along closed circuit paths \cite{Berry1}, a choice he did not explicitly explain.

As Dirac argued, monopole charges are determined by the properties of strings \cite{Dirac}, but these string charges cannot be directly inferred from density zeros. Monopole charges are closely related to the degeneracy degree of the energy spectrum and are also influenced by the dispersion form around degeneracy points. For a degeneracy point with linear dispersion, an N-fold degeneracy point yields a maximum monopole charge of \(\mu_{\text{max}}=(N-1)/2\), while the monopole charges of other eigenstates are integer-spaced values lying within the range \([-(N-1)/2, (N-1)/2]\) \cite{Berry1}. The string charge $m$ (defined as the phase variation around the string divided by \(2\pi\)) can be determined from the monopole charge as \(m=2\mu\) \cite{Dirac}. For instance, the aforementioned double degeneracy point corresponds to a monopole charge of \(\mu=\pm 1/2\). With knowledge of the monopole and string charges, phase variations for each eigenstate can be predicted using Eq.~\eqref{eq15}. However, a systematic account of monopole charges for arbitrary energy spectrum forms remains to be developed, which will be addressed in future work.

When the charges of strings and monopoles are unknown, their information can still be derived by investigating phase variations through the time-dependent Schr\"{o}dinger equation under the quantum adiabatic theorem. While this process resembles Berry's approach \cite{Berry1}, it is rooted in the original Diracian definition of topological vector potentials and effective magnetic fields \cite{Dirac}.

\section{Berry connection and curvature derived by Dirac monopole theory}
The non-integrable phase should varies with changing parameters $(X,Y,Z)$.  We can investigate the phase evolution of $|V_{\pm}(\mathbf{R})\rangle e^{i \gamma_{\pm}(\mathbf{R})}$ with varying parameters $\mathbf{R}$, which means that $\mathbf{R}$ depends on $t$ in the this case.
The phase evolution can be derived from the time-dependent  Schr\"{o}dinger equation
\begin{eqnarray}\label{timedep}
i \frac{d } {d t} \psi = H(\mathbf{R}) \psi.
\end{eqnarray}
We consider the initial state of $\psi$ is one eigensate $|V_{\pm}(\mathbf{R_c})\rangle e^{i \gamma_{\pm}(\mathbf{R_c})}$, where the $\mathbf{R_c}$ is one arbitrary position in the parameter space. The phase $\gamma_{\pm} (\mathbf{R_c})$ can be chosen as a reference point to define the relative phase distribution in the whole parameter space. Then the initial state can be written as $\psi(t=0)=|V_{\pm}(\mathbf{R_c})\rangle$. The state at an arbitrary time $t$ can be marked as $|V_{\pm}(\mathbf{R})\rangle e^{i \gamma_{\pm}(\mathbf{R})} e^{i \phi (t)}$, if the instantaneous eigen equation $H(\mathbf{R})|V_{\pm}(\mathbf{R})\rangle e^{i \gamma_{\pm}(\mathbf{R})}=E_{\pm}(\mathbf{R})|V_{\pm}(\mathbf{R})\rangle e^{i \gamma_{\pm}(\mathbf{R})} $ (the quantum adiabatic theorem) holds well \cite{Berry1}.  By substituting $|V_{\pm}(\mathbf{R})\rangle e^{i \gamma_{\pm}(\mathbf{R})} e^{i \phi (t)}$ to Eq.~\eqref{timedep}, we have $
i \frac{d |V_{\pm}(\mathbf{R})\rangle } {d \mathbf{R}} \frac{d \mathbf{R}}{d t}- |V_{\pm}(\mathbf{R})\rangle \frac{d \gamma_{\pm}(\mathbf{R})}{d \mathbf{R}} \frac{d \mathbf{R}}{d t}-|V_{\pm}(\mathbf{R})\rangle \frac{d \phi(t)}{d t}
=  E_{\pm} (\mathbf{R}) |V_{\pm}(\mathbf{R})\rangle$. Considering the adiabatic condition $\frac{d \mathbf{R}}{d t}\ll 1$, we can investigate them at different orders of them \cite{JLiu2010}. The zero order terms give the usual dynamical phase $\phi(t)=-\int_0^t E_{\pm}(\mathbf{R}) d t$. The first order terms give us that $|V_{\pm}(\mathbf{R})\rangle \frac{d \gamma_{\pm}(\mathbf{R})}{d \mathbf{R}}   =  i \frac{d |V_{\pm}(\mathbf{R})\rangle } {d \mathbf{R}}$. Then  we can obtain the expression for the non-integrable phase factor,
\begin{eqnarray}\label{timedep4}
	\gamma_{\pm}(\mathbf{R})  = \int_{\mathbf{R}_c}^{\mathbf{R}} \frac{i \langle V_{\pm}(\mathbf{R})| \nabla_{\mathbf{R}} V_{\pm}(\mathbf{R})\rangle } {\langle V_{\pm}(\mathbf{R})|V_{\pm}(\mathbf{R})\rangle } \cdot d \mathbf{R}.
\end{eqnarray}
This means that we can know the relative phase distribution of $\gamma_{\pm} (\mathbf{R})$ in the whole parameter space through varying parameters slowly. From the existence of above the Dirac strings and its endpoints, we can qualitatively know that the phase variations along arbitrary closed pathes are generally non-zero, which brings the relative phase distribution of $\gamma_{\pm}(\mathbf{R})$ depends on the integration pathes. This argument can be checked directly by calculating the phase variation along two different pathes between $\mathbf{R}_c $ and $\mathbf{R}$. The calculation results are helpful for understanding the striking characters of non-integrable phase factor given by Dirac \cite{Dirac}. This also means that the $\gamma_{\pm}(\mathbf{R})$ can not be an integrable function of $\mathbf{R}$. As done by Dirac in the real space, the explicit vector potential can be defined  by the phase gradient in the parameter space, which is
\begin{eqnarray}
\tilde{\mathbf{A}}(\mathbf{R}) = \nabla_{\mathbf{R}}  \gamma_{\pm}(\mathbf{R})= \frac{i \langle V_{\pm}(\mathbf{R})| \nabla_{\mathbf{R}} V_{\pm}(\mathbf{R})\rangle } {\langle V_{\pm}(\mathbf{R})|V_{\pm}(\mathbf{R})\rangle }.
\end{eqnarray}
The existence of Dirac strings with endpoints ensure that the vector potential admit topological singularities. If there are no endpoints for the strings in the parameter space, the vector potentials can be still defined but become trivial. The endpoints of strings in the above two-mode model brings nontrivial vector potentials, e.g., the three components of the vector potential for $E_+$ branch can be calculated as
$\tilde{A}_x=\frac{Y}{2 R (R+Z)}+\frac{i X (2 R+Z)}{2 R^2 (R+Z)},
\tilde{A}_y=-\frac{X}{2 R (R+Z)}+\frac{i Y (2 R+Z)}{2 R^2 (R+Z)},
\tilde{A}_z=\frac{i (R+Z)^2}{2 R^2 (R+Z)}$.
It should be noted that the $|V_{n}(\mathbf{R})\rangle$ is not normalized, which makes the the vector potential has imaginary parts. The curl of imaginary parts of the vector potential $\tilde{\mathbf{A}}$ is always zero, and they can be ignored safely. Therefore, the physical vector potential for $E_+$ branch is
\begin{eqnarray}
\mathbf{A} (\mathbf{R}) = \frac{Y \mathbf{e}_X- X \mathbf{e}_Y}{2 R (R+Z)}.
\end{eqnarray}
The monopole charge for $|V_{\pm}(\mathbf{R})\rangle$ are indeed $\mp \frac{1}{2}$. The vector potential is the well-known Berry connection \cite{Berry1}. Its curl gives the Berry curvature, which corresponds to effective magnetic fields of a Dirac monopole.

I derive the vector potential and effective magnetic field from the eigenstates, following the original Dirac monopole theory with the aid of adiabatic theory. In contrast, Berry derived the curvature expression with involving the energy spectrums, with solving the difficulties of choosing single-valued eigenstates in the parameter space \cite{Berry1}. In fact, the locally single-valued basis for an eigenstate is easily chosen without taking the non-integrable phase factor, and the chosen eigenstate's form do not vary the physical effects. Namely, the choice of eigenstate forms just modify the location of strings, but has no effects on the endpoints. Therefore there is no need to choose the eigenstate forms with any special limitations.

Berry further pointed out that the singularities of curvature always located at the degeneration points of energy spectrums \cite{Berry1}, based on the curvature expression with involving the energy spectrums.  Namely, the monopole-like fields are uncovered by the calculated Berry curvature, which can not be shown by the Berry connections. The analysis on energy degeneracy could be used to predict the monopole existence qualitatively before calculating Berry curvature in details. In contrast, the monopole fields can be predicted directly and qualitatively from the Dirac string properties before calculating vector potentials, for Dirac's picture.  Clearly, the endpoints of string precisely correspond to the degeneration points for the Hermitian systems. But our recent analysis on non-Hermitian systems indicate that the endpoints of string do not necessarily  correspond to the degeneration points \cite{YuZhao}. I think the interplay between the Dirac string and energy degenerations should be discussed further, which should induce some new monopole types.
\begin{figure}[t]
\begin{center}
\includegraphics[width=\linewidth]{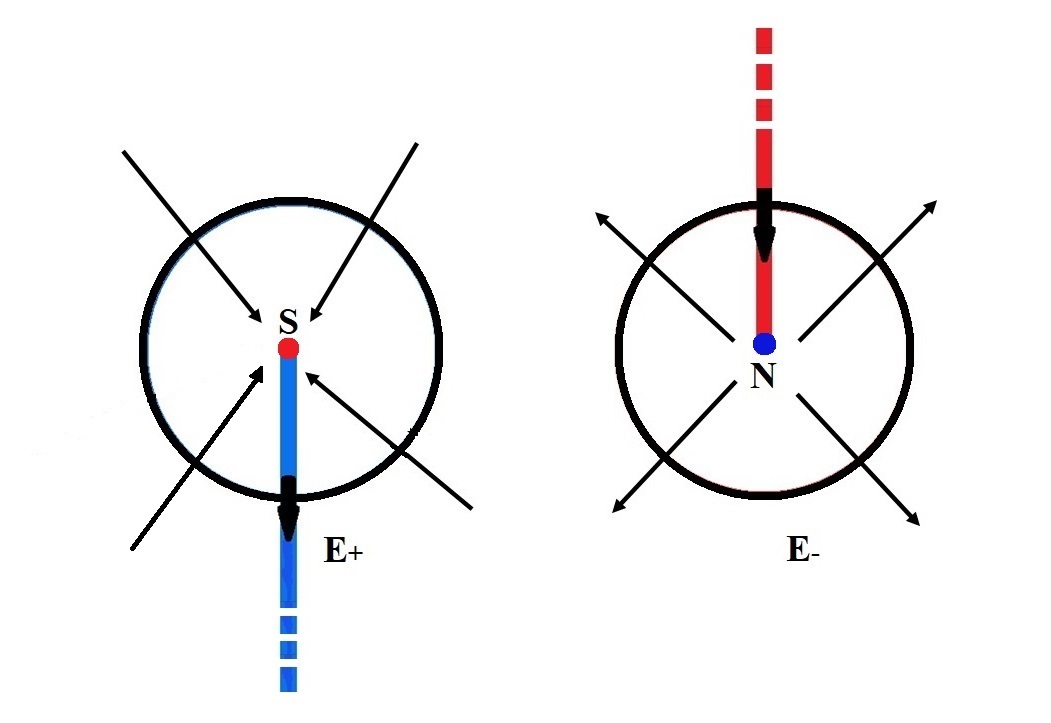}
\end{center}
\caption{It is illustrated that the eigenstates of the $E_{\pm}(\mathbf{R})$ branches admit  inverse charges at the endpoints of Dirac strings, and the magnetic fluxes of the two strings are also inverse. The monopole admits a $\mp 1/2$ charge for the   $E_{\pm}(\mathbf{R})$ branch, and the $2\pi$ magnetic flux directs outside (inside) the unit sphere for the $E_{+}(\mathbf{R})$ ($E_{-}(\mathbf{R})$) branch.}\label{Fig2}
\end{figure}

\section{Geometric phases given by visualizing Dirac strings with endpoints}
With a reference point \(\mathbf{R}_c\) and a specified path protocol, the phase \(\gamma_{\pm}\) across the entire parameter space can be calculated via the integral: \(\gamma_{\pm}(\mathbf{R}) = \int_{\mathbf{R}_c}^{\mathbf{R}} \frac{Y dX - X dY}{2 R (R + Z)}\).
For instance, choosing \(\mathbf{R}_c = (0, -1, 0)\) and a path protocol \((0, -1, 0) \rightarrow (X, -1, 0) \rightarrow (X, Y, 0) \rightarrow (X, Y, Z)\) (with each segment being a straight line) allows computation of the phase distribution \(\gamma_{+}(\mathbf{R})\). Distinct path protocols yield different phase distributions, a direct consequence of the non-integrable nature of the phase. This characteristic explains why Berry focused on phase variations along closed circuits \cite{Berry1}; in Dirac's framework, circuit paths are similarly natural for analyzing geometric phases, as reflected in Eq.~\eqref{eq15}.

Dirac originally described phase variations around diverse points (e.g., circuits enclosing Dirac strings, their endpoints, or other points) using an abstract wavefunction \cite{Dirac}. The phase variation given by \(\oint_C \frac{Y dX - X dY}{2 R (R + Z_j)}\) helps illuminate Dirac's discussions on total phase changes around large closed curves. For the \(E_+\) branch, Eq.~\eqref{eq15} can also be used to calculate phase variations for curves in the hemisphere (Fig.~\ref{Fig1}) once the monopole charge is known; here, we demonstrate this explicitly using Eq.~\eqref{timedep4}.
We calculated phase variations for five distinct circuits on the unit sphere (\(X^2 + Y^2 + Z^2 = 1\)), as shown in Fig.~\ref{Fig1}. These circuits correspond to \(Z_j = \pm 0.98, \pm 0.5, 0\), with each circuit defined by \(X^2 + Y^2 = 1 - Z_j^2\). For simplicity, \(\mathbf{R}_c\) was chosen on each circuit, and all circuits are traversed counterclockwise when viewed from the positive Z-axis. Our calculations using \(\oint_C \frac{Y dX - X dY}{2 R (R + Z_j)}\) yield the following phase variations for the five circuits in the hemisphere (Fig. \ref{Fig1}):
For \(Z_j = 0\) (circuit \(X^2 + Y^2 = 1\) with \(\mathbf{R}_c = (1, 0, 0)\)), the phase variation is \(\Delta \gamma_+ = -\pi\). For \(Z_j = 0.98\) and \(Z_j = 0.5\), the phase variations are \(-0.02\pi\) and \(-\pi/2\), respectively. For \(Z_j = -0.98\) and \(Z_j = -0.5\), the phase variations are approximately \(-1.98\pi\) and \(-3\pi/2\), respectively.
\begin{figure}[b]
	\begin{center}
		\includegraphics[width=\linewidth]{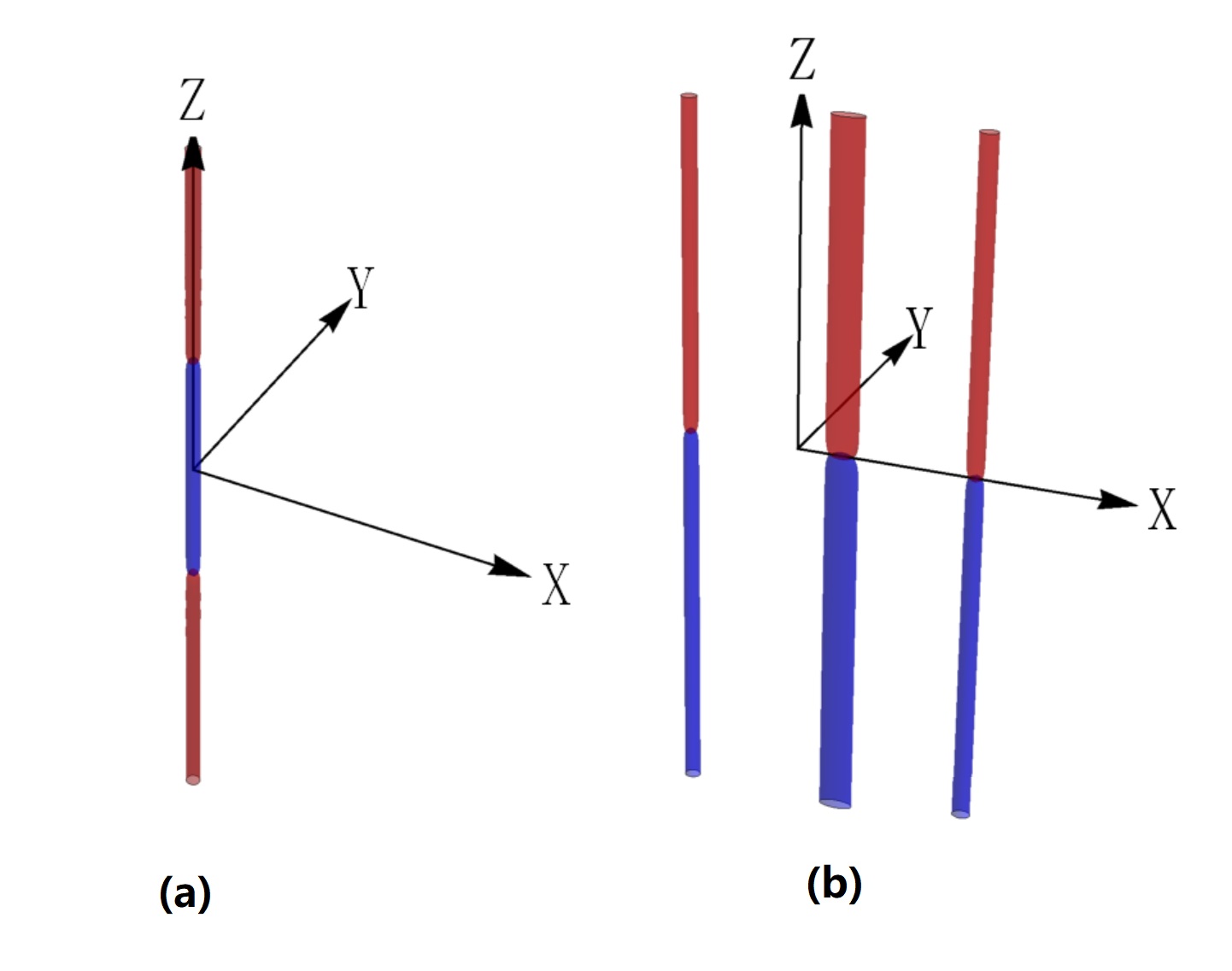}
		\end{center}
		\caption{The Dirac string and endpoints structures for the eigenstates of the Hamilton Eq.~\eqref{Hamiltonian} with replacing $Z$ by $Z^2-Z_0^2$ for (a) or $X$ by $(X-X_1)(X-X_2) (X-X_3)$ for (b). The blue lines and red lines denote the contour surface (0.001) of eigenstates' modulus for $E_{+}(\mathbf{R})$ and $E_{-}(\mathbf{R})$ branches, respectively. The parameters are chosen as $Z_0=0.5$, $X_1=-0.5$, $X_2=0.2$, and $X_3=0.8$.}\label{Fig3}
\end{figure}

The distinct phase variations between the upper and lower hemispheres arise from the presence of the Dirac string. For a counterclockwise circuit approaching \(Z_j \rightarrow -1\), the phase variation approaches \(-2\pi\), indicating a magnetic flux of \(2\pi\) from the string, directed outward from the unit sphere. This corresponds to the effective magnetic field for the \(E_+\) branch shown in the left panel of Fig.~\ref{Fig2}. For the \(E_-\) branch, the monopole charge is \(+1/2\); small closed circuits around its Dirac string exhibit a \(2\pi\) flux directed inward, with the effective magnetic field shown in the right panel of Fig.~\ref{Fig2}. Notably, different eigenstates possess distinct monopole charges---a feature absent in Dirac's original real-space monopole theory \cite{Dirac}.
Using the visualized Dirac strings with endpoints, phase variations for curves in the \(E_+\) hemisphere (Fig.~\ref{Fig1}) can be more conveniently computed via
Eq.~\eqref{eq15} by evaluating the solid angles of the surfaces bounded by the five curves. The Dirac string corresponds to the singular line of the vector potential \cite{Peres}; while the singular string is often considered unobservable, theoretical work has proposed monopoles without Dirac strings \cite{Wu-Yang,Brandt}.

For paths passing through the degeneracy point (the endpoint of the Dirac string), defining magnetic flux is non-trivial \cite{dmm}, but phase variations can still be computed via \(\Delta \gamma_{\pm}(\mathbf{R}) = \int_{\mathbf{R}_c}^{\mathbf{R}} \frac{Y dX - X dY}{2 R (R + Z)}\). For example, choosing \(\mathbf{R}_c = (-1, 0, 0)\) and the path \((-1, 0, 0) \rightarrow (X, 0, 0) \rightarrow (1, 0, 0)\) yields:
\begin{eqnarray}
\Delta\gamma_{+}&=&\int_{-1}^{1} \lim\limits_{Y\rightarrow0^{\pm}} \frac{Y}{2 (X^2+Y^2)}dX\nonumber\\
&=&\int_{-1}^{1} \frac{\pm \pi \delta(X)}{2 }dX=\pm \pi/2.
\end{eqnarray}

More complex scenarios with multiple nodal lines and endpoints can be considered by modifying the Hamiltonian in Eq.~\eqref{Hamiltonian}---e.g., replacing $Z$ with \(Z^2 - Z_0^2\) or $X$ with \((X - X_1)(X - X_2)(X - X_3)\). As shown in Fig.~\ref{Fig3}, these modifications yield:
Two monopoles at \((0, 0, \pm Z_0)\) for the \(E_{\pm}(\mathbf{R})\) branches (Fig.~\ref{Fig3}(a)). Three monopoles at \((X_{1,2,3}, 0, 0)\) (Fig.~\ref{Fig3}(b)).
Phase variations around arbitrary circles in the parameter space can be directly analyzed using Dirac's framework, given known monopole charges.

Notably, while degeneracy points always correspond to wavefunction zeros, wavefunction zeros do not necessarily coincide with degeneracy points. These observations aid in distinguishing wave dislocations \cite{wavedislo, Karjanto} from topological band theory \cite{band, band2}. Our recent analysis of non-Hermitian systems reveals that Dirac string endpoints no longer correspond to degeneracy points \cite{YuZhao}, suggesting that the interplay between Dirac strings and energy degeneracies requires further exploration---potentially leading to new classes of monopoles.

\section{Conclusion and discussions}
I for the first time extend the original Dirac monopole theory from real space to the parameter space, and explicitly demonstrate that the well-known Berry connection and Berry curvature can be rigorously derived from the non-integrable phase factors induced by the endpoints of nodal lines (Dirac strings) following Dirac's theoretical framework. It is clearly shown that in Hermitian systems, the endpoints of Dirac strings correspond exactly to the accidental degeneracy points of energy eigenvalues. A striking feature distinct from Dirac monopoles in real space \cite{Dirac} is revealed: different eigenstates in the parameter space possess distinct monopole charges. These results fundamentally indicate that the Berry geometric phase is essentially the non-integrable phase factor induced by Dirac strings with endpoints in the parameter space, thereby establishing a unified theoretical framework for understanding Dirac monopole theory and geometric phases. Our research differs fundamentally from previous studies that explored Dirac-type singularities or monopole-like characteristics in Berry phases by focusing on phase variations around level crossing points \cite{Bhandari,Fujikawa1,Fujikawa2}. Those studies discussed Dirac strings and monopoles based on calculated Berry connections and curvatures, whereas ours is rooted in the original Dirac monopole theory with a primary focus on the zeros of wave functions.

I further propose a more effective and unified approach to investigate monopoles and geometric phases for arbitrary Hamiltonians. For linear Hermitian systems (while non-Hermitian and nonlinear systems require further exploration due to the lack of a general monopole theory formalism), the approach proceeds in three steps: first, calculate the eigenvalues and eigenstates of the Hamiltonian; second, analyze the nodal line properties of eigenstates without normalization (a key step, as normalization obscures density zeros). According to Dirac's monopole picture, the existence of Dirac strings with endpoints can qualitatively predict the presence of monopole fields in the parameter space and momentum space. If no such endpoints exist, it implies the absence of observable geometric phases or other topological effects, rendering the calculation of vector potentials (usually called Berry connections) unnecessary. Third, when Dirac strings with endpoints are identified, derive topological vector potentials from non-integrable phase factors following Dirac's method---these potentials fully quantify the topological or geometric properties.

In contrast, the conventional approach based on Berry's framework requires analyzing energy degeneracies using the full energy spectrum to infer monopole-like Berry curvatures at accidental degeneracy points, which inherently demands information beyond a single eigenvalue. A critical advantage of our Dirac-based approach is that it allows determining the presence of monopole-like fields solely through nodal line analysis of an arbitrary eigenstate, without relying on information from other eigenstates. This makes it particularly efficient for complex systems where obtaining the full energy spectrum is challenging. Notably, our analysis of non-Hermitian systems reveals that the endpoints of Dirac strings do not necessarily correspond to degeneracy points, suggesting that the interplay between Dirac strings and energy degeneracies warrants further investigation, which is expected to lead to new types of monopoles \cite{YuZhao}.

The Dirac monopole theory thus exhibits greater generality than the Berry geometric phase framework. For instance, while some studies have observed wave functions with Dirac strings and endpoints in real space, uncovering Dirac monopoles in synthetic magnetic fields---systems where Berry's framework is nearly inapplicable---our recent work shows that Dirac's analytical formalism can reveal hidden monopoles in extended complex planes \cite{Zhaoliu1,Zhaoliu2}, a task intractable with Berry's framework. Our results not only deepen the understanding of the relationship between Berry geometric phases and the original Dirac monopole theory but also provide a strong impetus for exploring geometric phases through the implementation and extension of Dirac monopole theory.

\section*{Acknowledgments}
The author wishes to express his gratitude to Professor Jie Liu for his insightful and helpful suggestions about this work. The author is grateful to Dr. Bin Sun and Liang Duan for their helpful discussions.  This work was supported by the National Natural Science Foundation of China (Contract No. 12375005, 12235007, 12247103).

\end{document}